\documentstyle[12pt]{article}
\begin{document}
\topmargin 0pt
\oddsidemargin 5mm
\headheight 0pt
\headsep 0pt
\topskip 5mm

\begin{flushright}
NBI-HE-93-36\\
\hfill
August 1993
\end{flushright}

\begin{center}
\hspace{10cm}

\vspace{15pt}
{ \bf
THE ROKHSAR-KIVELSON MODEL OF QUANTUM DIMERS\\
AS A GAS OF FREE FERMIONIC STRINGS }
\end{center}

\vspace{15pt}

\begin{center}

Peter Orland\raisebox{.6ex}{*}\footnote{\raisebox{.6ex}{*}
Work
supported by PSC-CUNY
Research Award Program grant nos. 662467 and 663368.}\\
\vspace{25pt}

{The Niels Bohr Institute, \\
Blegdamsvej 17, DK-2100, \\
Copenhagen {\O}, Denmark \\
orland@nbivax.nbi.dk} \\

\vspace{15pt}

and\\

\vspace{15pt}

The City University of New York,\\
Baruch College,\\
17 Lexington Ave.,\\
New York, NY 10010, U.S.A.\\
orlbb@cunyvm.cuny.edu\raisebox{.6ex}{\dag}
\footnote{\raisebox{.6ex}{\dag}Permanent address}\\

\vspace{25pt}

{\bf Abstract}
\end{center}

{\em The $2+1$-dimensional quantum dimer model on a square
lattice, proposed by Rokhsar and Kivelson as a theory of layered
superconductivity, is shown to be equivalent to a many-body theory
of free, transversely oscillating strings obeying Fermi
statistics. A Jordan-Wigner construction for string field
operators is presented. Topological
defects are shown to be linearly confined in
pairs by dynamical strings. Exact upper and lower bounds
are placed on the ground-state
energy and the string tension. It is argued that the system is in a
spin-fluid phase and that there is no gap in the
excitation spectrum.}

\vfill

\newpage

\section{Introduction}

In the last few years, progress has been made in exactly solving
statistical and quantum models in a total of three dimensions. Some
cases of the six-edge statistical model
in three dimensions\raisebox{.6ex}{1}, the $U(1)$ lattice
gauge magnet\raisebox{.6ex}{2} and the quantum dimer model
on a hexagonal lattice\raisebox{.6ex}{3} in $2+1$ dimensions
have been solved
exactly.

In this article, the Rokhsar-Kivelson
model (RKM) on the square lattice\raisebox{.6ex}{4}
will studied from a similiar
point of view. This system is a quantum dimer model on a square
lattice, proposed as an effective theory of layered
superconductors. It is found that, like the systems solved in
references 1,2,3, the RKM is a system of non-interacting Fermionic
strings. While the one-string problem is not, unfortunately, exactly
solvable, it can be understood physically without much effort.  More
general models with a ``diagonal"
coupling or with dynamical
holons\raisebox{.6ex}{4} are not
accessible to the techniques used
here.

It is found here that the behaviour of the RKM without
the ``diagonal" term is qualitatively
similiar to that of the quantum dimer model on the
hexagonal lattice; topological
defects (holons) are confined into pairs, but there is no gap in
the spectrum and no long-range order. The ground
state can be thus be thought of
as a spin fluid. It can be shown that dimer
configurations are the ground states of a particular Heisenberg
model invented by Klein\raisebox{.6ex}{5}. Presumably introducing further
interactions in this Heisenberg model gives a system whose
low-lying states are described by the RKM.

It is perhaps appropriate at this point to review some of the
history and issues surrounding the RKM and
related problems. Rokhsar and Kivelson's model
has two parameters; a ``resonance" coupling $J$ and a
``diagonal" coupling $V$. These are coefficients of two different
terms in the Hamiltonian. They were able to solve the model
at $V=J$ where they found that the ground state was a superposition
of every possible dimer state. For
$V>J$ they showed that the ground
state was a ``spin-staggered" state with spontaneous breaking
of rotational and translation invariance. They suggested that for
$V<J$ a ``spin-liquid" state with no long-range order should
appear. They also argued that a
``valence-bond-solid" or ``column" phase should appear
for $V<0$. In this phase so-called column
configurations dominate, leading
to a four-fold
degeneracy of the ground state.Later, Dombr\'{e} and Kotliar, who
examined mean field theory for the Hubbard model
and Read and Sachdev\raisebox{.6ex}{6}, who studied
$SU(N)$ generalizations of square-lattice
Heisenberg antiferromagnets formally similiar
to the RKM, did not find
a spin-liquid phase, but did find a valence-bond-solid
phase appeared\raisebox{.6ex}{7}. The RKM
model was then examined directly by Fradkin
and Kivelson\raisebox{.6ex}{8} who used dilute gas methods for
the $V=0$ case and Sachdev, who
studied the same problem
numerically\raisebox{.6ex}{9}. Fradkin and Kivelson
argued that the $V=0$ ground state was a valence-bond-solid
state. Sachdev
found that the susceptibility of the order
parameter for valence-bond-solid formation was divergent
with the volume, and therby suggested that such a solid formed
for all $V<J$. Read and Sachdev later extended their analysis to
antiferromagnets
on hexagonal lattices and argued that valence-bond-solid formation
took place there as well\raisebox{.6ex}{10}. For at least
some Heisenberg models, the ground state is a
valence-bond solid\raisebox{.6ex}{11}. In
reference 3, it was shown that
a hexagonal-lattice quantum dimer model was in a gapless
fluid phase\raisebox{.6ex}{12}.

It will be shown here
that the ground state of the square-lattice
RKM at $V=0$ is
a fluid and not a solid. The
absence of long-range order is consistent with
Rokhsar and Kivelson's original picture of superconductivity. There
do appear to be power-law correlations which may explain Sachdev's
results\raisebox{.6ex}{9}.

In this paper, the string formulation is arrived at through the
use of a generalization of the Jordan-Wigner transformation
due to Dotsenko and Polyakov for the three-dimensional
Ising model\raisebox{.6ex}{13}. In previous
solutions\raisebox{.6ex}{1,2,3}
antisymmetrized string wave functions were used to diagonalize
the Hamiltonian. The main
reason for presenting
the more formal Jordan-Wigner construction here is that some
readers may find it more straightforward as well as more
interesting from a mathematical
point of view.

The RKM will first be converted to a triangular-lattice problem which
resembles the neutral $U(1)$ lattice
gauge magnet\raisebox{.6ex}{2,14,15}. Next, the Hilbert space
will be mapped into that of strings with an infinitely
short-range, infinitely repulsive interaction. The interaction is removed
by quantizing the strings as Fermions. The eigenfunctions and eigenvalues
of the RKM
can thereby be written in terms of the one-string eigenvectors
and eigenvalues. The one-string problem is
not exactly solvable, but physical arguments indicate
its low-lying spectrum is that of a $1+1$-dimensional massless
quantum field theory of Dirac particles. It is argued that the spectrum
of the full RKM is gapless. A
linear potential exists between topological defects, due to the
formation of dynamical strings. Exact upper and
lower bounds are found on both the ground-state energy and
the string tension.

\section{The Rokhsar-Kivelson Model}

The sites of a two-dimensional lattice will be
labeled by pairs of integers $(y^{1},y^{2})=\bf y$ and
bonds $({\bf y}, i)$ connecting $\bf y$ to
${\bf y} +{\hat{\bf i}}$. A spin state
$|s({\bf y},i)>$ is defined on each bond, by $s({\bf y}, i)=1/2$ when
a dimer
is present at $({\bf y}, i)$ and $s({\bf y}, i)=-1/2$, when no dimer
is present at $({\bf y}, i)$. A dimer
represents a $\pi$-valence
bond between two nearest-neighbor
copper atoms\raisebox{.6ex}{16}. The
Hamiltonian of the RKM is\raisebox{.6ex}{4}
\begin{equation}
H= J \,\sum_{\bf y} \,\sum_{\pm} \;
\sigma^{\pm}({\bf y}, 1)\sigma^{\mp}({\bf y} +{\hat {\bf 1}}, 2)
\sigma^{\pm}({\bf y} +{\hat {\bf 2}}, 1)
\sigma^{\mp}({\bf y}, 2)\;. \label{1.1}
\end{equation}
Here $\sigma^{\pm}({\bf y}, i)=(1/2)(\sigma^{x}({\bf y},i) \pm
i\sigma^{y}({\bf y},i))$, where
$\sigma^{x}$, $\sigma^{y}$ and $\sigma^{z}$ are the
usual Pauli matrices. The
Hamiltonian density applied to a
plaquette (square) changes that plaquettes's
state according to the rule:
\\
\vspace{2pt}
\begin{center}

\begin{picture}(85,15)(0,0)
\put(0,0){\line(1,0){15}}
\put(0,15){\line(1,0){15}}

\put(55,0){\line(0,1){15}}
\put(70,0){\line(0,1){15}}

\put(30,4){$\rightleftharpoons$}

\end{picture}
\\
\vspace{10pt}
\end{center}
with all other states annihilated. The RKM is a $U(1)$ lattice gauge
theory\raisebox{.6ex}{2,14,15}; $H$
commutes with the Gauss' law operator defined
at a site $\bf y$, adjacent to four links $l$
\begin{equation}
G({\bf y})=\sum_{l} \sigma^{z}(l)  \;. \label{1.2}
\end{equation}
Dimer states $|\Psi>$ satisfy
\begin{equation}
[G({\bf y})+2] |\Psi>=0  \;.  \label{1.3}
\end{equation}
The Hamiltonian is identical to that of the model discussed
in reference 2, though the Gauss' law condition (\ref{1.2}) is
different. The gauge-invariant Wilson loop operator
defined on a closed contour of consecutive
links
$C=\{l_{1}, l_{2},...,l_{P}\}$ is
\begin{equation}
A(C)=\sigma^{+}(l_{1})\sigma^{-}(l_{2})\sigma^{+}(l_{3})\;...
\;\sigma^{-}(l_{P}) \;.  \label{1.4}
\end{equation}
A topological defect is a site $\vec y$ at
which no dimers are attached. Such defects
are characterized by
\begin{equation}
[G({\bf y})+4] |\Psi>=0  \;.  \label{1.5}
\end{equation}
In reference 4 a second ``diagonal" term was added to the Hamiltonian, namely
\begin{equation}
H_{d}= -V\sum_{\bf y} {\cal S}^{\pm}({\bf y}, 1)
{\cal S}^{\mp}({\bf y} +{\hat {\bf 1}}, 2)
{\cal S}^{\pm}({\bf y} +{\hat {\bf 2}}, 1)
{\cal S}^{\mp}({\bf y}, 2)\;, \label{1.6}
\end{equation}
where ${\cal S}^{\pm}$ is defined by
\begin{equation}
{\cal S}^{\pm}({\bf y}, i)
=\frac{1\pm \sigma^{z}({\bf y}, i)}{2}\;. \label{1.7}
\end{equation}
In this paper, only the $V=0$ case will be considered.

\section{Reduction to a Triangular Lattice}

The RKM can be reduced to a spin model
on a triangular lattice. The triangular lattice
will be drawn as a square lattice with extra
diagonal bonds. Upon making this reduction, $90^{o}$ rotational
invariance is no longer manifest. Consider
the bonds $({\bf y},2)$, with $y^{1}+y^{2}$ odd. These
are half the bonds parallel to ${\hat 2}$. The
occupation state at each of these bonds is a redundant
degree of freedom. All such bonds may be contracted to
points, leaving a triangular lattice. To
see this, consider the possible dimer configurations
around one such bond:\\
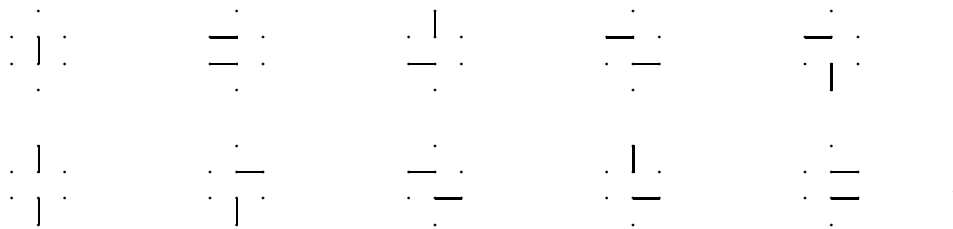
\vspace{2pt}
\begin{center}
\begin{picture}(405,50)(0,0)

\multiput(0,10)(75,0){5}{\circle*{.3}}
\multiput(0,20)(75,0){5}{\circle*{.3}}
\multiput(10,0)(75,0){5}{\circle*{.3}}
\multiput(10,10)(75,0){5}{\circle*{.3}}
\multiput(10,20)(75,0){5}{\circle*{.3}}
\multiput(10,30)(75,0){5}{\circle*{.3}}
\multiput(20,10)(75,0){5}{\circle*{.3}}
\multiput(20,20)(75,0){5}{\circle*{.3}}

\put(10,10){\line(0,1){10}}
%
%
%
%

\put(75,10){\line(1,0){10}}
\put(75,20){\line(1,0){10}}

%
%
%
%
%
%

\put(150,10){\line(1,0){10}}
\put(160,20){\line(0,1){10}}

%
%
%
%

\put(225,20){\line(1,0){10}}
\put(235,10){\line(1,0){10}}

%
%
%
%
%
%

\put(300,20){\line(1,0){10}}
\put(310,0){\line(0,1){10}}

%
%
%
%

\end{picture}\\
\begin{picture}(405,50)(0,0)

\multiput(0,10)(75,0){5}{\circle*{.3}}
\multiput(0,20)(75,0){5}{\circle*{.3}}
\multiput(10,0)(75,0){5}{\circle*{.3}}
\multiput(10,10)(75,0){5}{\circle*{.3}}
\multiput(10,20)(75,0){5}{\circle*{.3}}
\multiput(10,30)(75,0){5}{\circle*{.3}}
\multiput(20,10)(75,0){5}{\circle*{.3}}
\multiput(20,20)(75,0){5}{\circle*{.3}}

\put(10,20){\line(0,1){10}}
\put(10,0){\line(0,1){10}}

%
%

\put(85,20){\line(1,0){10}}
\put(85,0){\line(0,1){10}}

%
%
%
%

\put(150,20){\line(1,0){10}}
\put(160,10){\line(1,0){10}}

%
%
%
%
%
%

\put(235,10){\line(1,0){10}}
\put(235,20){\line(0,1){10}}

%
%
%
%

\put(310,10){\line(1,0){10}}
\put(310,20){\line(1,0){10}}

%
%
%
%
%
%

\put(355,12){.}

\end{picture}\\
\end{center}
\vspace{10pt}
The middle bond is now contracted to a point (Fig.1).  The
sites of the
triangular lattice will be labeled by
${\bf x}=(x_{1},x_{2})$ and the bonds connecting ${\bf x}$ to
${\bf x}+{\hat {\bf 1}}$, connecting ${\bf x}$ to ${\bf x}+{\hat {\bf 2}}$
and connecting ${\bf x}$ to $\bf x -{\hat {\bf 1}} -{\hat {\bf 2}}$
by $({\bf x},1)$, $({\bf x},2)$ and $({\bf x},3)$ respectively. The
state at each bond
is represented by a spin $|s({\bf x},i)>$. Next apply
the unitary transformation which flips the spins
at all horizontal bonds $\prod_{\bf x} \sigma^{x}(\bf x, 1)$
(i.e. occupied bonds at $({\bf x},1)$
are now replaced by empty bonds and {\em vice-versa}). The possible
configurations at a site of the triangular lattice are
now:\\
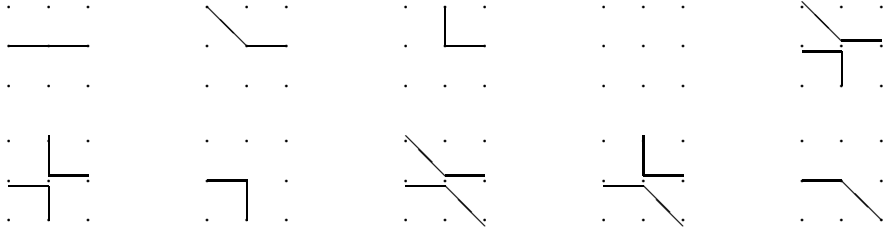
\vspace{2pt}
\begin{center}
\begin{picture}(405,50)(0,0)

\multiput(0,0)(75,0){5}{\circle*{.3}}
\multiput(0,15)(75,0){5}{\circle*{.3}}
\multiput(0,30)(75,0){5}{\circle*{.3}}
\multiput(15,0)(75,0){5}{\circle*{.3}}
\multiput(15,15)(75,0){5}{\circle*{.3}}
\multiput(15,30)(75,0){5}{\circle*{.3}}
\multiput(30,0)(75,0){5}{\circle*{.3}}
\multiput(30,15)(75,0){5}{\circle*{.3}}
\multiput(30,30)(75,0){5}{\circle*{.3}}

\put(0,15){\line(1,0){30}}

%
%
%
%

\put(75,30){\line(1,-1){15}}
\put(90,15){\line(1,0){15}}

%
%

\put(165,15){\line(1,0){15}}
\put(165,15){\line(0,1){15}}

%
%

%
%
%
%

\put(300,32){\line(1,-1){15}}
\put(315,0){\line(0,1){13}}
\put(300,13){\line(1,0){15}}
\put(315,17){\line(1,0){15}}


\end{picture}\\
\begin{picture}(405,50)(0,0)

\multiput(0,0)(75,0){5}{\circle*{.3}}
\multiput(0,15)(75,0){5}{\circle*{.3}}
\multiput(0,30)(75,0){5}{\circle*{.3}}
\multiput(15,0)(75,0){5}{\circle*{.3}}
\multiput(15,15)(75,0){5}{\circle*{.3}}
\multiput(15,30)(75,0){5}{\circle*{.3}}
\multiput(30,0)(75,0){5}{\circle*{.3}}
\multiput(30,15)(75,0){5}{\circle*{.3}}
\multiput(30,30)(75,0){5}{\circle*{.3}}

\put(0,13){\line(1,0){15}}
\put(15,17){\line(1,0){15}}
\put(15,0){\line(0,1){13}}
\put(15,17){\line(0,1){15}}


\put(75,15){\line(1,0){15}}
\put(90,0){\line(0,1){15}}

%
%

\put(150,13){\line(1,0){15}}
\put(165,17){\line(1,0){15}}
\put(165,13){\line(1,-1){15}}
\put(150,32){\line(1,-1){15}}


\put(225,13){\line(1,0){15}}
\put(240,13){\line(1,-1){15}}

\put(240,17){\line(1,0){15}}
\put(240,17){\line(0,1){15}}


\put(300,15){\line(1,0){15}}
\put(315,15){\line(1,-1){15}}

%
%

\end{picture}\\
\end{center}
\vspace{10pt}
respectively. The fifth, sixth, eighth and ninth
configurations have purposely been
drawn with a break. The new Hilbert space
is restricted by the Gauss law
\begin{equation}
g({\bf x)}|\Psi>=0  \;,   \label{2.1}
\end{equation}
where
\begin{equation}
g({\bf x})= \sigma^{z}({\bf x},1)-\sigma^{z}({\bf x},2)+\sigma^{z}({\bf x},3)
-\sigma^{z}({\bf x}-{\hat {\bf 1}},1)+\sigma^{z}({\bf x}-{\hat{\bf 2}},2)
-\sigma^{z}({\bf x}+{\hat {\bf 1}}+{\hat {\bf 2}},3) \; . \label{2.2}
\end{equation}

The Hamiltonian
is now
\begin{eqnarray}
H&=&\sum_{{\bf x}}\,\sum_{\pm}
     [\sigma^{\pm}({\bf x},1) \sigma^{\pm}({\bf x},2)
     \sigma^{\mp}({\bf x}+{\hat{\bf 2}},3)   \nonumber \\
 &+&\sigma^{\pm}({\bf x}+{\hat{\bf 2}},3)
\sigma^{\mp}({\bf x}+{\hat{\bf 2}},1)
\sigma^{\mp}({\bf x}+{\hat{\bf 1}},2)]  \;\;, \label{2.6}
\end{eqnarray}
This $H$ commutes with $g(\bf x)$ defined in
(\ref{2.2}). It interchanges
states on triangles according to:
\\
\vspace{2pt}
\begin{center}
\begin{picture}(155,15)(0,0)
\put(0,0){\circle*{.3}}
\put(15,0){\circle*{.3}}
\put(0,15){\circle*{.3}}
\put(0,0){\line(0,1){15}}
\put(0,0){\line(1,0){15}}
\put(45,0){\circle*{.3}}
\put(60,0){\circle*{.3}}
\put(45,15){\circle*{.3}}
\put(45,15){\line(1,-1){15}}
\put(25,3){$\rightleftharpoons$}
\put(70,0){,}

\put(100,15){\circle*{.3}}
\put(100,0){\circle*{.3}}
\put(85,15){\circle*{.3}}
\put(85,15){\line(1,-1){15}}
\put(145,0){\circle*{.3}}
\put(145,15){\circle*{.3}}
\put(130,15){\circle*{.3}}
\put(130,15){\line(1,0){15}}
\put(145,0){\line(0,1){15}}
\put(110,3){$\rightleftharpoons$}


\put(152,0){.}
\end{picture}
\end{center}
\vspace{10pt}

A typical configuration of the square
lattice is shown in Fig. 2. The
basis states of the Hilbert space are
strings extending across the lattice. Two strings
never overlap on any link.

The strings are not arbitrary paths. Any
unit segment of string on a vertical bond (that is, a bond
in the $2$-direction) or on a bond at a $45^{o}$ (that is, a
bond of string parallel to ${\hat{\bf 1}} -{\hat{\bf 2}}$)
must be attached at the top
to a segment of string on a horizontal bond (in the
$1$-direction), coming from the left, unless the vertical
bond is on the boundary. In addition, a segment
of string on a vertical or $45^{o}$ bond
must be attached at the bottom to a segment of string on a horizontal
bond, going off to the right. Three impossible configurations for two
adjacent segments of a single string are:\\
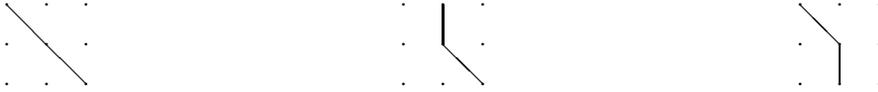
\vspace{2pt}
\begin{center}
\begin{picture}(405,35)(0,0)

\multiput(0,0)(150,0){3}{\circle*{.3}}
\multiput(0,15)(150,0){3}{\circle*{.3}}
\multiput(0,30)(150,0){3}{\circle*{.3}}
\multiput(15,0)(150,0){3}{\circle*{.3}}
\multiput(15,15)(150,0){3}{\circle*{.3}}
\multiput(15,30)(150,0){3}{\circle*{.3}}
\multiput(30,0)(150,0){3}{\circle*{.3}}
\multiput(30,15)(150,0){3}{\circle*{.3}}
\multiput(30,30)(150,0){3}{\circle*{.3}}

\put(0,30){\line(1,-1){30}}


\put(165,15){\line(1,-1){15}}
\put(165,15){\line(0,1){15}}


\put(315,0){\line(0,1){15}}
\put(300,30){\line(1,-1){15}}


\end{picture}
\end{center}
\vspace{10pt}

The strings interact though
a short-range, infinitely strong repulsive interaction
at links. Systems of Bosonic strings with
such ``hard-core" repulsion in 2+1 dimensions
are equivalent to systems of free Fermionic strings\raisebox{.6ex}{2,3}.

Assume the original lattice of the RKM has the shape of a rectangle of
vertical dimension $2N$ and horizontal dimension $L$, where
$L$ is odd. Select for contraction those links $({\bf y},2)$
for which $y^{1}+y^{2}$ is an odd number. Then the
resulting triangular lattice has the appearance shown at the bottom
of Fig.1. It
is possible to see that all but one site of the left boundary $x^{1}=1$ of the
triangular lattice is an endpoint of exactly one string. There
are therefore $N$ strings attached to this boundary (it contains
$N+1$ sites). The site which is not
an endpoint of a string is
that at the bottom of the left boundary (it is possible for
a string to pass though this site, but not end there). Similiarly,
all but one site of the right boundary $x^{1}=L$ of the
triangular lattice is an endpoint of exactly one string, the
exception being the site at the top of the right boundary. The
number of strings is conserved
by the Hamiltonian. With this choice
of boundary condition, the top and bottom boundaries place no restriction
on the shape a string may take.

If the system were in a valence-bond-solid phase, the
ground state would be spontaneously
broken. It would be dominated by one of the four column
configurations shown in Fig.3. On the triangular
lattice the column configurations become those shown in Fig.4. It
will be argued here, that such symmetry breaking does not
take place  (provided there is no diagonal term (\ref{1.6})
included in the Hamiltonian). The
argument hinges on the string spectrum being gapless. In
fact, a much simpler argument can
be made against valence-bond-solid formation. Consider
any plaquette containing two
dimers in a column configuration. The state
of this plaquette is completely free to fluctuate. There
is no preference for the two dimers on this plaquette to be aligned along
the $1$-direction or the $2$-direction. The
configuration of each of the other plaquettes containing
two dimers can also fluctuate. The solid cannot maintain its
integrity and must melt.

\section{The One-String Problem}

Each
string is made of segments attached end-to-end. A string configuration
is described by a sequence of numbers $X(k)$, each an integer or
half-integer, defined below. The
midpoint of each segment has $1$-coordinate $x^{1}=n/2+1/2$, $n$ even, while
the endpoints have $1$-coordinates $x^{1}=n/2+1/2$, $n$
odd. In this way, both
midpoints and endpoints of segments are labeled by the
horizontal coordinate $n$. The number
$X(k)$ is an integer or half-integer equal to the value of
$x^{2}$ at the midpoint of the segment of string at
$x^{1}=k+1/2$. Clearly
$n=2k$. The eigenstates of the one-string Hamiltonian in the
Schr\"{o}dinger picture are wave functionals
$S[X]=S[X(1),..., X(k),...,X(L)]$. Though the links $(\bf x, 1)$
with $x^{1}=L$ do not exist on the triangular lattice, the number
$X(L)$ is still needed. The purpose of $X(L)$ is
to specify how the string
ends at the right boundary. These
string wave functionals vanish for any choice of
$[X]$ which does not correspond to a string. Define the
raising and lowering operators on this space
of wave-functionals by
\begin{equation}
[\Delta(k),X(k')]=\delta_{k, \;k'}\;,\;\;
[\Delta^{\dag}(k),X(k')]=-\delta_{k,\; k'} \; . \label{3.2}
\end{equation}

The string configuration is next mapped into a
one-dimensional spin configuration $\{s(n)\}$, $s(n)=\pm 1$. A
spin at odd $n$
is down unless the string segment at $x^{1}=n/2$ is parallel
to $\hat{\bf 2}$. A
spins at even $n$ is down, unless the string segment
whose midpoint is $x^{1}=n/2$ is tilted at $45^{o}$, i.e. parallel to
$\hat{ \bf 1}-\hat{ \bf 2}$. The one-string Hamiltonian is equivalent to a
spin-chain Hamiltonian
under an inner-product-preserving transformation $A$:
\begin{equation}
{\cal H}=A\,h\,A^{\dag} \;,   \label{3.3}
\end{equation}
defined by
\begin{eqnarray}
A\,\sigma^{z}(2l-1)\,A^{\dag} &=& 2{\delta}_{X(l),\;X(l-1)-1} -1 \;,
\nonumber \\
  A\,\sigma^{z}(2l)\,A^{\dag} &=& 2{\delta}_{X(l+1),\;X(l)-1/2}
                                  \;{\delta}_{X(l),\;X(l-1)-1/2}
\;-1 \;, \nonumber \\
A\,\sigma^{+}(2l-1)\,A^{\dag} &=& \prod_{m=2l}^{2L+1}
\Delta  (m)^{2}\;,                  \nonumber \\
  A\,\sigma^{+}(2l)\,A^{\dag} &=& \Delta (2l)\;\prod_{m=2l+1}^{2L+1}
\Delta (m)^{2}\;,                 \nonumber \\
A\,\sigma^{-}(2l-1)\,A^{\dag} &=& \prod_{m=2l}^{2L+1}
\Delta ^{\dag}(m)^{2}             \;,\nonumber \\
  A\,\sigma^{-}(2l)\,A^{\dag} &=&  \Delta ^{\dag}(2l)\;\prod_{m=2l+1}^{2L+1}
{\Delta ^{\dag}(m)}^{2} \;.
\label{3.4}
\end{eqnarray}

Certain restrictions must be placed on spin
configurations if they are to be identified by (\ref{3.4})
with string configurations. They
are:\begin{enumerate}
\item the spins $s(n)$ and $s(n+1)$ cannot both be $+1$,
\item the spins $s(2l)$ and $s(2l+2)$ cannot both be $+1$,
\end{enumerate}
The spin-chain Hamiltonian which describes the string is
\begin{equation}
h=J\sum_{n=1}^{2L }\sigma^{\pm}_{n}\sigma^{\mp}_{n+1}
+\lambda \sum_{n=1}^{2L } (1+\sigma^{z}_{n})(1+\sigma^{z}_{n+1})
+\lambda \sum_{l=1}^{L-1}
(1+\sigma^{z}_{2l})(1+\sigma^{z}_{2l+2})\;,  \label{3.5}
\end{equation}
where $\lambda$ is taken to infinity to enforce the above
restrictions 1 and 2. Notice that the last term
breaks the invariance under translation by one lattice spacing (this
invariance is also broken by the boundary conditions). The
number of up spins is fixed by virtue of the fact that the
spin chain has a global $U(1)$ invariance:
\begin{equation}
[\;h,\sum_{n=1}^{2L+1} \sigma^{z}_{n}\;]\;=\;0 \;\;.\label{3.6}
\end{equation}
The
Hamiltonian (\ref{3.5})
describes spinless Fermions which can hop from site to
site, with strong nearest-neighbor and next-to-nearest-neighbor
repulsion. The Fermion fields are
constructed through the Jordan-Wigner
transformation\raisebox{.6ex}{17}:
\begin{equation}
\psi^{\dagger}_{n}=\sigma^{+}_{n}\prod_{m<n}\sigma^{z}_{m}\;,\;\;
\psi_{n}=\sigma^{-}_{n}\prod_{m<n}\sigma^{z}_{m}\;.\label{3.7}
\end{equation}
These obey $[\psi^{\dagger}_{n},\psi_{m}]_{+}=\delta_{n,m}$, with all other
anticommutators vanishing. The filling
fraction must be $1/4$ by virtue
of the boundary conditions (this is consistent
with (\ref{3.6}) ). Fermions seperated from each other
by more than two lattice spacings
have the usual hopping
dispersion relation $E_{k}=2J\cos k$, hence a Fermi
sea forms in the ground state of (\ref{3.5}). If
the second term in (\ref{3.5})
is ignored, the low-lying excitations are
one-dimensional massless Dirac particles\raisebox{.6ex}{17}. It
will be argued that this is true for the system (\ref{3.5}) as well.

The
additional repulsive interactions in (\ref{3.5}) should not
produce a gap above the Fermi sea. There are two reasons why this
should be so. The first of these is that a repulsive
short-range interaction cannot form bound
states in the Fermi fluid. Secondly, the spin-chain is
parity-invariant (in the sense that a spin at a given site is coupled
in the same fashion to its left neighbor as its right neighbor), so
that the left and right Dirac sectors should be
uncoupled, which is inconsistent with a gap. However, I do not have a
proof of this claim.

The
Hamiltonian (\ref{3.5}) cannot
be solved exactly, but it should be possible to study its
spectrum by numerical methods. It is possible, however, to find exact
upper and lower bounds for its ground-state energy. While
this can actually be done for the finite open chain, I will do so
only for the ground-state energy per unit
length in the thermodynamic limit.

For simplicity, assume that the Hamiltonian (\ref{3.5} is put on
lattice of length ${\cal L}$ with periodic boundary conditions, i.e.
$\sigma^{i}_{n+{\cal L}}=\sigma^{i}_{n}$. Consider
the Hamiltonian
\begin{equation}
h_{R}=J\sum_{n=1}^{{\cal L}}\sigma^{\pm}_{n}\sigma^{\mp}_{n+1}
+\lambda \sum_{n=1}^{{\cal L}}\sum_{j=1}^{R}
(1+\sigma^{z}_{n})(1+\sigma^{z}_{n+j}) \;,  \label{3.8}
\end{equation}
on such a lattice and where, as before, $\lambda$ is
taken to infinity. The number $R$ is the range of the potential
and can be any non-negative integer (The $R=0$ case
is the XX chain\raisebox{.6ex}{17}). For
a specified filling fraction $f={\cal N}/{\cal L}$, the
ground-state energy per site of $h_{1}$ and $h_{2}$
will be shown to be lower and upper bounds, respectively, of the
ground-state energy per site of $h$. In addition the ground-state
energy per site of $h_{R}$ can be explicitly calculated for
any filling fraction. This model was considered recently
by G\'{o}mez-Santos, who noted that for
particular fillings
its solution is a Luttinger liquid\raisebox{.6ex}{18} .

It is straightforward
to find all the eigenstates and eigenvalues of $h_{R}$. Consider
the one-up-spin eigenstates. These states are labeled by wave number
$k=2\pi p/{\cal L}$ where $p=0,...,{\cal L}-1$ :
\begin{equation}
|k>= \sum_{n=1}^{{\cal L}} e^{ikn} \sigma_{n}^{+} |\Omega> \;, \label{3.9}
\end{equation}
where $|\Omega>$ is the state with all spins down (zero
filling fraction). These states are eigenstates of both $h$ and
$h_{R}$ with eigenvalues $E_{k}=2\cos k$. The eigenstates of $h$ and
$h_{R}$ must
have the
form
\begin{eqnarray}
|k_{1},...,k_{\cal N}>
&= (\sum_{n})'\;
          \exp\,\,(ik_{1}n_{1}+\cdot \cdot \cdot
          +ik_{\cal N}n_{\cal N})     \nonumber \\
&\times  \Psi_{k}(n_{1},...,n_{\cal N})
\;\sigma_{n_{1}}^{+} \cdot \cdot \cdot \sigma_{n_{\cal N}}^{+}
\;|\Omega> \;, \label{3.10}
\end{eqnarray}
where the sum is over $1 \leq n_{1}<n_{2}< \cdot \cdot \cdot
<n_{\cal N} \leq {\cal L}$. A state of the form (\ref{3.10}) is an eigenstate
of $h_{R}$ if and only if
it is an eigenstate of the XX model (free Fermions) and satisfies
the condition that $\Psi_{k}(n_{1},...,n_{{\cal N}})=0$
if for some $r$, $n_{r+1}-n_{r} \le R$. Thus any eigenstate of $h_{R+1}$ is
also an eigenstate of $h_{R}$ with the same eigenvalue. A state
of the form (\ref{3.10}) is an eigenstate
of $h$ if and only if
it is an eigenstate of the $XX$ model and satisfies
the condition that
$\Psi_{k}(n_{1},...,n_{{\cal N}})=0$
if either : \begin{enumerate}
\item $n_{r+1}-n_{r} \le 1$, for any $r$,
\item $n_{2p+1}-n_{2p} \le 2$, for any $p$.
\end{enumerate} These conditions imply that
any eigenstate of $h_{2}$ is also an eigenstate of $h$ and
any eigenstate of $h$ is also an eigenstate of $h_{1}$. Therefore
the ground state at filling fraction $f$ of $h_{2}$ is an eigenstate
of $h$ and the ground state at filling fraction $f$ of $h$ is an eigenstate
of $h_{1}$. If $\epsilon_{R}(f)$ is the ground-state energy per site
of $h_{R}$ and $\epsilon(f)$ is the ground-state energy per site
of $h$, then the above considerations imply
\begin{equation}
\epsilon_{1}(f) \le \epsilon(f) \le \epsilon_{2}(f) \;. \label{3.11}
\end{equation}

The simplest way to calculate $\epsilon_{R}(f)$ is to notice
that $h_{R}$ is just a free Fermion problem (XX chain) on an effective
lattice
of size ${\cal L}'={\cal L}-{\cal N}R={\cal L}(1-fR)$. The
effective filling fraction is
\begin{equation}
f'=\frac{{\cal N}}{{\cal L}'}=\frac{f}{1-fR} \;. \label{3.12}
\end{equation}
Now ${\cal L}'\epsilon_{0}(f')$ is by definition, the ground-state energy
of the free Fermion problem on a lattice of size ${\cal L}'$ and
filling fraction $f'$. The ground-state energy per site of $h_{R}$ at
filling fraction $f$ must therefore be
\begin{equation}
\epsilon_{R}(f)=\frac{{\cal L}'\epsilon_{0}(f')}{{\cal L}}
= (1-fR)\epsilon_{0}(\frac{f}{1-fR})  \;. \label{3.13}
\end{equation}
The ground-state energy density the free
Fermi problem on a large lattice
with this filling fraction is
\begin{equation}
\epsilon_{0}(\frac{f}{1-fR})=-\frac{2J}{\pi}
\sin(\frac{\pi f}{1-fR})\;. \label{3.14}
\end{equation}
In the thermodynamic limit, the energy per site of the string
on a lattice on dimensions $N\times L$ is given by
$\mu=\epsilon(1/4) /2$. From
(\ref{3.13}) and (\ref{3.14}) the inequality (\ref{3.11}) becomes
\begin{equation}
-\frac{1}{\sqrt{3} \pi}\;J \le \mu
\le -\frac{1}{\pi}\;J  \;, \label{3.15}
\end{equation}
or
\begin{equation}
-0.184\;J \le \mu \le -0.159\;J  \;. \label{3.16}
\end{equation}

\section{The $N$-String Problem}

The solution of the $N$-string problem is essentially  the
same as in references
2 and 3. First the one-string Hamiltonian is found. Then
the product of $N$ one-string eigenfuntionals is antisymmetrized
in each $X(n)$. This
gives an eigenfunctional of ${\cal H}_{1}+...+{\cal H}_{N}$, where
${\cal H}_{k}$ is the Hamiltonian for the $k^{th}$ string. It is
in fact an eigenfunctional of the full Hamiltonian, since
it is guaranteed to vanish whenever strings overlap.

Antisymmetrization can ``break up" some of the
strings; it leads to choices of $[X]$ which
are not continuously connected strings. The
wave functionals automatically vanish whenever this is the case.

Let ${\bf S}_{N}$ be the permutation group of $N$ objects. Given
$N$ strings, there are $N$ values of $X(k)$ for each $k$, which
will be labeled $X^{1}(k),...,X^{N}(k)$. Permutations
$s_{k} \varepsilon {\bf S}_{N}$ will
act by sending the values of $X(k)$ to
$X^{s_{1}(1)}(k),...,X^{s_{k}(N)}(k)$. The
antisymmetrized product of $N$ string wave functionals quantum numbers (which
are the
momenta of Fermions in each string) $\{\theta \}_{j}$ for $j=1,...,N$ is
\begin{eqnarray}
S_{\{\theta\}_{1}...\{\theta\}_{N}}[X]=
[\prod_{k=1}^{L} \frac{1}{N!} \sum_{s_{k} \varepsilon {\bf S}_{N}}
\;sgn(s_{k})\;]\;\prod_{j=1}^{N}
S_{\{\theta \}_{j} }[X^{s_{1}(j)}(1),...,X^{s_{L}(j)}(L)] \;.\label{4.1}
\end{eqnarray}
The argument $[X]$ is then restricted to a fundamental
region ${\cal F}$ in which there is no overlapping of strings:
\begin{equation}
X^{1}(k)<
X^{2}(k)< ...
<X^{N}(k)  \;.  \label{4.2}
\end{equation}
This proceedure is consistent because
$S_{\{\theta\}_{1}...\{\theta\}_{N}}[X]$ vanishes just outside
${\cal F}$, where
links from different strings touch. With this
restriction, ${\cal H}_{1}+...+{\cal H}_{N}$ can be
identified with ${\cal H}_{s}$, and
hence with the RKM Hamiltonian and
$S_{\{\theta\}_{1}...\{\theta\}_{N}}[X]$ are indeed the
correct eigenstates. The energy eigenvalues are
\begin{equation}
{\cal E}_{\{\theta\}_{1},...,\{\theta\}_{N} } =
E_{\{\theta \}_{1}}+...+E_{\{\theta \}_{N}}    \;. \label{4.3}
\end{equation}
The reader who is not entirely
convinced of the results
of this section can find a more complete
discussion of many-string theory in reference 3.

\section{String Field Theory}

A more formal method of converting the RKM
to a many-string problem is the string-Jordan-Wigner
transformation, similiar to
that discussed for the three-dimensional Ising
model\raisebox{.6ex}{13}.

Let $[X]=[X(1),..., X(k),...,X(L)]$ be
a set of links $l_{1}, l_{2},...$ connecting the left
boundary to the right boundary. These links do
not neccesarily constitute a
string configuration consistent with the
rules discussed in Section 3, but are
assumed to be a connected path, such that, for each $k=1,...,L$ there is a
unique $X(k)$. String
field operators will be defined for each $[X]$.

A contour $[X]$ is shown in Fig.5. Let the
set of links of the
form $({\bf x}, 1)$, $({\bf x}, 3)$, but not
$({\bf x}, 2)$ {\em below} $[X]$ on
the triangular lattice be called
$\cal D$. The links of $\cal D$ are shown as dotted
lines in Fig.5. The string-destruction operator at $[X]$ is
\begin{equation}
\Psi [X]= [\prod_{l \varepsilon [X]} \sigma^{-}(l)]\;
[\prod_{l \varepsilon {\cal D}} \sigma^{z}(l)]
\;=\;[\sigma^{-}(l_{1})\sigma^{-}(l_{1}) \cdot \cdot \cdot ]
\;[\prod_{l \varepsilon {\cal D}} \sigma^{z}(l)]    \;. \label{4.4}
\end{equation}
Similiarly, the string-creation operator at $[X]$ is
\begin{equation}
\Psi^{\dag} [X]= [\prod_{l \varepsilon [X]} \sigma^{+}(l)]
\;[\prod_{l \varepsilon {\cal D}} \sigma^{z}(l)]
\;=\;[\sigma^{+}(l_{1})\sigma^{+}(l_{1}) \cdot \cdot \cdot ]
\;[\prod_{l \varepsilon {\cal D}} \sigma^{z}(l)]    \;. \label{4.5}
\end{equation}
The string fields satisfy certain anticommutation
relations. The $\Psi^{\dag}$'s anticommute
among themselves, as do the $\Psi$'s :
\begin{equation}
[\; \Psi^{\dag} [X]\, ,\; \Psi^{\dag} [Y]\;]_{+}=
[\Psi [X]\; ,\; \Psi [Y]\;]_{+}=0 \;,  \label{4.6}
\end{equation}
for any two contours $[X]$ and $[Y]$. A special case
of (\ref{4.6}) is
\begin{equation}
(\Psi^{\dag} [X])^{2}=
(\Psi [X])^{2}=0 \;.  \label{4.7}
\end{equation}
In addition to (\ref{4.6}), the string fields satisfy
\begin{equation}
[\;\Psi^{\dag} [X]\; ,\; \Psi [Y]\;]_{+}= \delta_{[X]\;[Y]} \;,  \label{4.8}
\end{equation}
Suppose $[X]$ and $[Y]$ are two contours which cross at the
points $\bf x$, $\bf z$, such that $[X]$ is everywhere
below $[Y]$ (Fig.6). By cutting both contours at
$\bf x$, $\bf z$, and then switching and re-attaching the
pieces two new contours, $[Z]$ and $[U]$, are obtained. The
string-field $\Psi^{\dag}$ operators satisfy
\begin{equation}
\Psi^{\dag} [X]\; \Psi ^{\dag} [Y]
=(-1)^{z^{2}-x^{2}}\;\Psi^{\dag} [Z]\; \Psi^{\dag} [U]  \;,  \label{4.9}
\end{equation}
Notice that if $[X]$ and $[Y]$ are string contours
satisfying the rules of Section 3, the contours $[Z]$ and $[U]$ will
not satisfy these rules.

States produced by applying products of operators
$\Psi^{\dag} [X]$ to the zero-string state $|Z.S.>$ (satifying
$\Psi [X]\;|Z.S.>=0$) are the second-quantized versions
of (\ref{4.1}).

For each $[X]$ define a vector $|[X]>$, such that $<[X]|[X]>=1$ and
if
$[X] \neq [Y]$, $<[X]|[Y]>=0$. These vectors
constitute an orthonormal
basis of a vector space $V$. The states which correspond
to the special string configurations
satisfying the rules of Section 3 span a subspace $W$. The
mapping $A$ defined in (\ref{3.3}) is a linear
transformation from $W$ to the spin-chain Hilbert space, which
preserves inner products. The
Hamiltonian (\ref{2.6}), with the constraint ({\ref{2.1}) enforced, is
\begin{eqnarray}
H&=&-\sum_{|[X]> \varepsilon W}\,\sum_{|[Y]> \varepsilon W}\;
     \Psi^{\dag} [Y]\;<[Y]|\;{\cal H}\;|[X]>\;\Psi [Y] \nonumber \\
 &=&-\sum_{|[X]> \varepsilon W}\,\sum_{|[Y]> \varepsilon W}\;
\Psi^{\dag} [Y]\;<[Y]|\;A\;h\;A^{\dag}\;|[X]>\;\Psi [Y] \;, \label{4.14}
\end{eqnarray}
where $h$ is the spin-chain Hamiltonian defined in equation
(\ref{3.5}).

The ground state is the
antisymmetrized product of ground-state string wave functionals, with
the coordinates
ordered as discussed previously. By (\ref{4.3}) and (\ref{3.15}), the
ground state $E_{0}$
energy  must satisfy
\begin{equation}
-\frac{1}{\sqrt{3} \pi}\;J \le \frac{E_{0}}{NL}
\le -\frac{1}{\pi}\;J  \;, \label{4.15}
\end{equation}
Correlations
in the ground state of an individual string have
only power law behaviour, as the
spectrum of the spin chain is relativistic
and gapless. Thus the ground state of the RKM has
power-law correlations, and is neither a
columnar nor a staggered state\raisebox{.6ex}{4}. An
excitation of the RKM
can be made by exciting a single string. Clearly
the gap is zero.

\section{Static Defects}

An important issue is how the system behaves
when topological defects are introduced. In models of
copper-oxide-layer superconductors, defects
are charged excitations and
their confinement implies Cooper pairing with a short
coherence length. Only static defects
will be discussed here. The square lattice is bipartite with
two geometrically
distinct kinds of sites, namely {\em red} sites with $x^{1}+x^{2}$ even
and {\em black} sites with $x^{1}+x^{2}$ odd. Since
the Hamiltonian (\ref{1.1})
commutes with the Gauss's law operator (\ref{1.2}), defect locations
are fixed. A red defect becomes a site where a string ends at the
right, whereas a black defect becomes a site where a
string begins at the
left. Free boundary conditions
for dimers on the square lattice
dictate imply that there
are an equal number of red and black defects.

An isolated defect on a red site
becomes one of the configurations \\
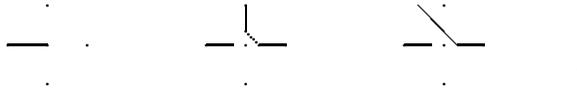
\vspace{2pt}
\begin{center}
\begin{picture}(405,35)(0,0)

\multiput(145,15)(75,0){3}{\circle*{.3}}
\multiput(115,15)(75,0){3}{\circle*{.3}}
\multiput(130,0)(75,0){3}{\circle*{.3}}
\multiput(130,15)(75,0){3}{\circle*{.3}}
\multiput(130,30)(75,0){3}{\circle*{.3}}

\put(115,15){\line(1,0){15}}

\put(190,15){\line(1,0){10}}
\put(210,15){\line(1,0){10}}
\put(205,20){\line(0,1){10}}
\multiput(205,20)(1,-1){6}{\circle*{.05}}

\put(265,15){\line(1,0){10}}
\put(270,30){\line(1,-1){15}}
\put(285,15){\line(1,0){10}}

\put(330,5){,}

\end{picture}
\end{center}
\vspace{10pt}
on the triangular lattice, while an isolated defect at a black
site becomes one of the configurations\\
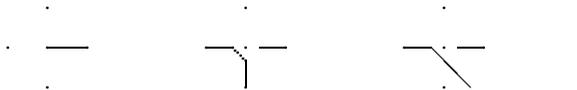
\vspace{2pt}
\begin{center}
\begin{picture}(405,35)(0,0)

\multiput(145,15)(75,0){3}{\circle*{.3}}
\multiput(115,15)(75,0){3}{\circle*{.3}}
\multiput(130,0)(75,0){3}{\circle*{.3}}
\multiput(130,15)(75,0){3}{\circle*{.3}}
\multiput(130,30)(75,0){3}{\circle*{.3}}

\put(130,15){\line(1,0){15}}

\put(190,15){\line(1,0){10}}
\put(210,15){\line(1,0){10}}
\put(205,0){\line(0,1){10}}
\multiput(200,15)(1,-1){6}{\circle*{.05}}

\put(265,15){\line(1,0){10}}
\put(285,15){\line(1,0){10}}
\put(275,15){\line(1,-1){15}}

\put(330,5){.}

\end{picture}
\end{center}
\vspace{10pt}

The red defects and black defects behave like positive and
negative charges, respectively in a confining $U(1)$ gauge
theory\raisebox{.6ex}{19}. If a string is removed
by placing a red defect at the left boundary
and a black defect is placed at the
right boundary, then the lowest energy state
is identical to the ground state with no defects, minus the energy
of one string. Since the individual strings have a negative
energy per unit length (because the filling fraction in the
spin chain is greater than zero), the defects are joined
by a physical ``string hole" with positive energy per unit length. This
energy is minus the ground-state energy
of one string. The red defect and the black defect
are thus bound by a constant attractive force through
the formation of this electric string. The string
tension is defined to be the asymptotic
energy per unit seperation on the original
square lattice. Since the seperation of the defects
on the square lattice
is $L$, the string tension is $\mu$ (which
satisfies the bounds (\ref{3.15}). The
system should be rotational-invariant at large distances (since
the gap is zero) so the string tension will only depend on the seperation
of two defects, if that seperation is large.

\section{Conclusions}

It has been shown that
the RKM is equivalent to a gas of transversely oscillating
``hard" strings. A qualitative analysis shows that
the energy spectrum is gapless, indicating a fluid phase. Defects
are confined by dynamical
strings. Further numerical work on the spin chain (\ref{3.5})
describing a single string should reveal further properties, such as
the precise values of the string tension and ground-state energy.

The physical excitations are probably closed
strings. A closed string excitation is produced by disturbing one of
the strings in the ground state (the string ``sea") along a finite
region of its length, then antisymmetrizing and ordering the
string coordinates.

It seems likely that for the range of diagonal coupling $J\ge V \ge 0$ there
is a fluid phase, with a phase transition to a valence-bond solid
for some negative
$V$.

\section*{Acknowledgements}

I would like to thank Nick Read and Subir Sachdev
for educating me
on a number of issues and for
many useful comments
on the manuscript. I am grateful to the the
Niels Bohr Institute staff for their hospitality
and the Danish Research Council for some financial support.

\vspace{20pt}

{\large \bf  References}
\vspace{10pt}

\begin{enumerate}

\item P. Orland, Int. J. Mod. Phys. {\bf B5}, 2401 (1991).
\item P. Orland, Nucl. Phys. {\bf B372}, 635 (1992).
\item P. Orland, Phys. Rev. {\bf B47}, 11280 (1993).
\item D. Rokhsar, S.A. Kivelson, Phys. Lett. {\bf 61}, 2376 (1988).
\item D.J. Klein, J. Phys. {\bf A15}, 661 (1982); J.T. Chayes, L.
Chayes and S.A. Kivelson, Commun. Math.
Phys. {\bf 123}, 53 (1989).
\item T. Dombre and G. Kotliar, Phys. Rev. {\bf B39}, 855
(1989); N. Read and S. Sachdev, Nucl. Phys. {\bf B316}, 609
(1989); Phys. Rev. Lett. {\bf 62}, 1694 (1989); Int.
J. Mod. Phys. {\bf B5}, 219 (1991).
\item It was incorrecly stated in reference 3 that the authors
of reference 6 had made similiar
conclusions concerning the RKM.
\item E. Fradkin and S.A. Kivelson, Mod. Phys. Lett. {\bf B4}, 225
(1990); E. Fradkin, {\bf Field Theories of Condensed
Matter Systems}, Addison-Wesley (1991).
\item S. Sachdev, Phys. Rev. {\bf B40}, 5204 (1989).
\item N. Read and S. Sachdev, Phys. Rev. {\bf B42}, 4568 (1990).
\item M.P Gelfand, R.R.P. Singh and D.A. Huse, Phys. Rev.
{\bf B40}, 10801 (1989); M.P. Gelfand, Phys.
Rev. {\bf B42}, 8206 (1990).
\item It
was pointed out to me by N. Read that the boundary
conditions taken in reference 3 are in fact incompatible
with global spin-solid formation. However, it can be
shown that the basic results of reference 3 are valid even
for boundary conditions not suffering from this
difficulty.
\item Vl.S. Dotsenko and A.M. Polyakov in
Advanced Studies in Pure Mathematics {\bf 16}, ed. M. Jimbo, T.
Miwa and A. Tsuchiya, Kinkuniya, Tokyo,
171 (1988); A.R. Kavalov and A. Sedrakyan, Nucl. Phys. {\bf B321},
682 (1989);  P. Orland, Phys. Rev. Lett. {\bf 59}, 2393 (1987).
\item D. Horn, Phys. Lett. {\bf 100B}, 149 (1981).
\item P. Orland and D. Rohrlich, Nucl. Phys. {\bf B338}, 647 (1990).
\item P.W. Anderson, Science {\bf235}, 1196 (1987).
\item Y. Nambu, Progr. Theoret. Phys. {\bf 5}, 1, E (1950); E.
Lieb, T. Schultz
and D. Mattis, Ann. Phys. {\bf 16}, 407 (1961).
\item G. G\'{o}mez-Santos, Phys. Rev. Lett. {\bf 70}, 3780 (1993).
\item A.M. Polyakov, Phys. Lett. {\bf 59}, 79 (1975); Nucl.
Phys. {\bf B120}, 429 (1977).

\end{enumerate}

\vfill
\newpage

\section*{Figure Captions}

\vspace{15pt}

\begin{itemize}

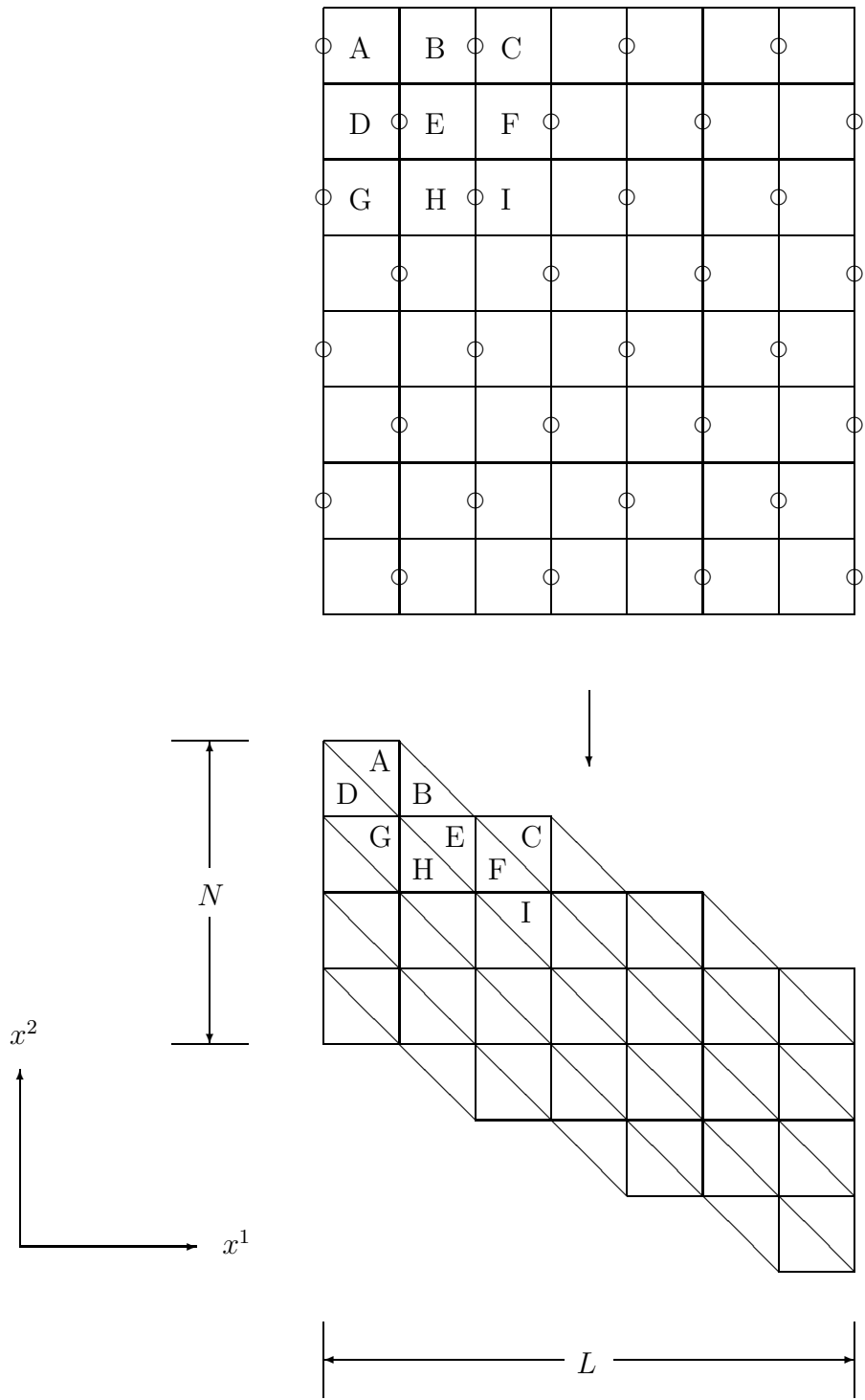
\item Figure 1 : Reduction of the square lattice to the triangular
lattice. The bonds marked with circles are eliminated, converting
each square (some of which are labeled by
labeled by A, B, C, etc.) to a triangle. For this particular
case $N=4$ and $L=7$.

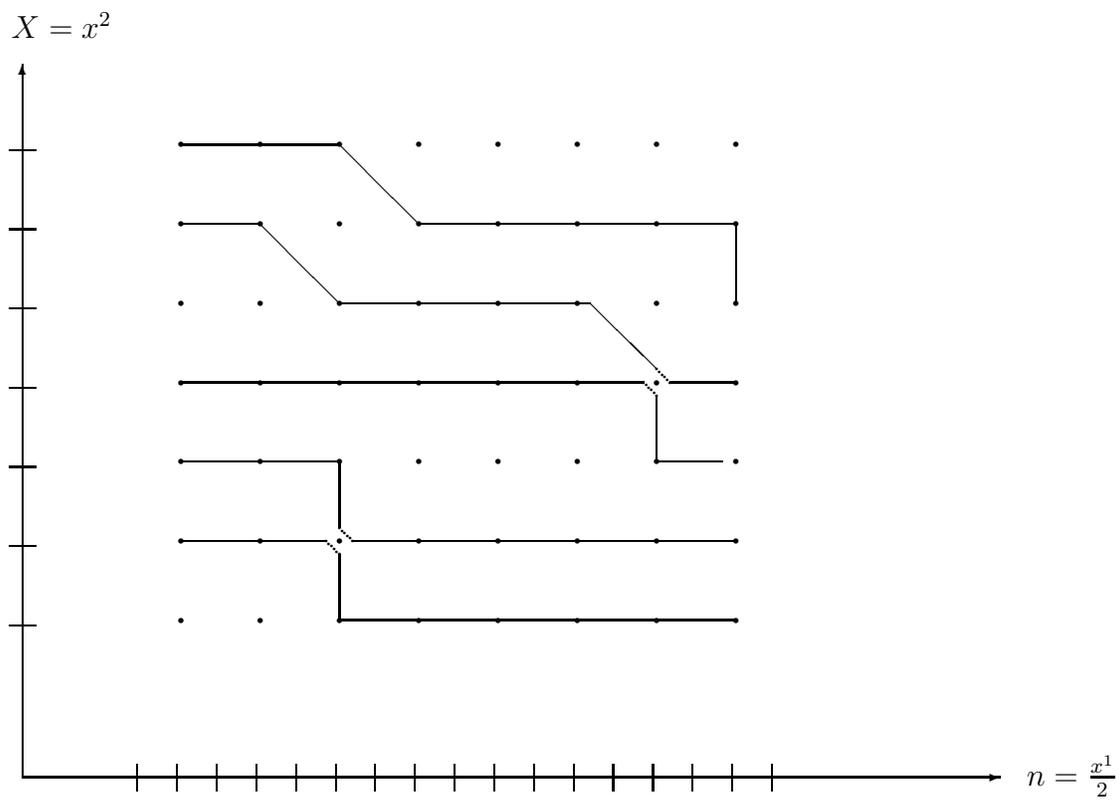
\item Figure 2 : A typical configuration of the
RKM, after reduction to the triangular lattice. Each dimer
configuration on the old lattice becomes a set of strings on
the new lattice.

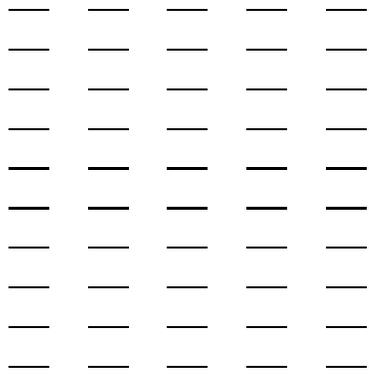
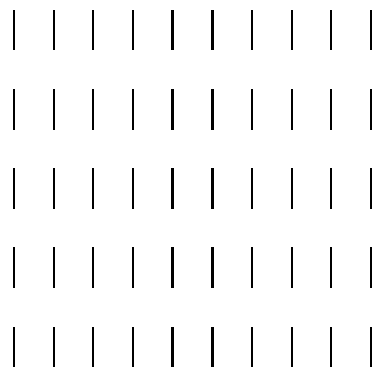
\item Figure 3 : Two of the
four basic four column configurations. By translating
(a) horizontally and (b) vertically
by one lattice spacing, the other two column
configurations are generated.

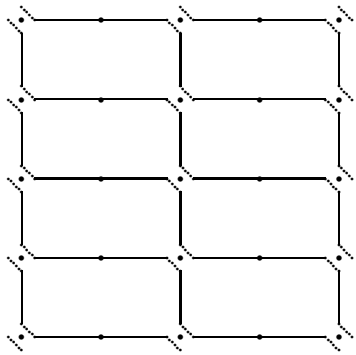
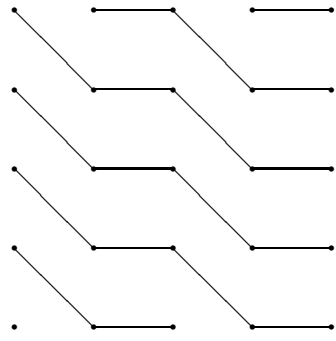
\item Figure 4 : The
configurations on the triangular lattice obtained
from the column configurations of Fig.3. The other two
such configurations are obtained by translating these by
one lattice spacing.

\item Figure 5 : How the Fermionic string fields $\Psi$ and $\Psi ^{\dag}$
are defined for a contour $[X]$. Notice that this particular choice
of $[X]$ is not a legitimate string configuration
according to the rules of Section 3. The dark lines indicate the
links of the string contour $[X]$, while the dotted lines
indicate the links in the set ${\cal D}$.

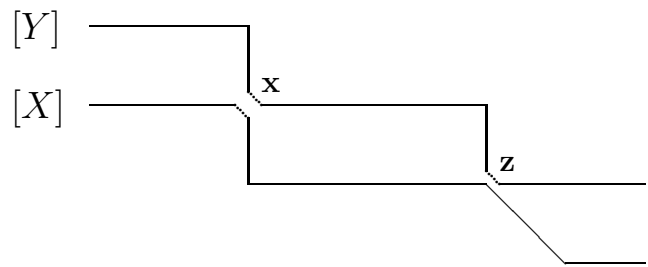
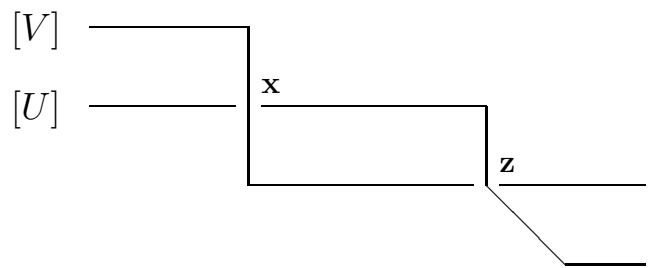
\item Figure 6 : The relationship between the contours
$[X]$, $[Y]$, $[Z]$ and  $[U]$.

\end{itemize}
\newpage


\begin{picture}(400,450)(0,0)

\put(-10,-70){\vector(1,0){70}}
\put(70,-73){$x^{1}$}

\put(-10,-70){\vector(0,1){70}}
\put(-14,10){$x^{2}$}

\multiput(110,180)(0,30){9}{\line(1,0){210}}

\multiput(110,180)(30,0){8}{\line(0,1){240}}

\multiput(110,225)(60,0){4}{\multiput(0,0)(0,60){4}{\circle{6}}}
\multiput(140,195)(60,0){4}{\multiput(0,0)(0,60){4}{\circle{6}}}

\put(120,400){A}
\put(150,400){B}
\put(180,400){C}
\put(120,370){D}
\put(150,370){E}
\put(180,370){F}
\put(120,340){G}
\put(150,340){H}
\put(180,340){I}

\put(215,150){\vector(0,-1){30}}

\put(110,40){\line(1,-1){60}}
\put(110,70){\line(1,-1){120}}
\put(110,100){\line(1,-1){180}}
\put(110,130){\line(1,-1){210}}
\put(140,130){\line(1,-1){180}}
\put(200,100){\line(1,-1){120}}
\put(260,70){\line(1,-1){60}}

\put(50,10){\line(1,0){30}}
\put(50,130){\line(1,0){30}}
\put(65,60){\vector(0,-1){50}}
\put(65,80){\vector(0,1){50}}
\put(60,65){$N$}

\put(290,-80){\line(1,0){30}}
\put(230,-50){\line(1,0){90}}
\put(170,-20){\line(1,0){150}}
\put(110,10){\line(1,0){210}}
\put(110,40){\line(1,0){210}}
\put(110,70){\line(1,0){150}}
\put(110,100){\line(1,0){90}}
\put(110,130){\line(1,0){30}}

\put(110,130){\line(0,-1){120}}
\put(140,130){\line(0,-1){120}}
\put(170,100){\line(0,-1){120}}
\put(200,100){\line(0,-1){120}}
\put(230,70){\line(0,-1){120}}
\put(260,70){\line(0,-1){120}}
\put(290,40){\line(0,-1){120}}
\put(320,40){\line(0,-1){120}}

\put(110,-130){\line(0,1){30}}
\put(320,-130){\line(0,1){30}}
\put(225,-115){\vector(1,0){95}}
\put(205,-115){\vector(-1,0){95}}
\put(210,-120){$L$}

\put(128,118){A}
\put(145,105){B}
\put(188,88){C}
\put(115,105){D}
\put(158,88){E}
\put(175,75){F}
\put(128,88){G}
\put(145,75){H}
\put(188,58){I}

\put(195,-170){Fig.1}

\end{picture}

\newpage


\begin{picture}(400,450)(0,0)

\put(0,200){
\multiput(60,124.5)(0,-30){7}{\multiput(0,15)(30,0){8}{\circle*{1.6}}}

\put(0,-210){
\put(60,349.5){\line(1,0){60}}
\put(120,349.5){\line(1,-1){30}}
\put(150,319.5){\line(1,0){120}}
\put(270,319.5){\line(0,-1){30}}

\put(60,319.5){\line(1,0){30}}
\put(90,319.5){\line(1,-1){30}}
\put(120,289.5){\line(1,0){95}}
\put(215,289.5){\line(1,-1){25}}
\multiput(240,264.5)(1,-1){6}{\circle*{.05}}
\put(245,259.5){\line(1,0){25}}

\put(60,259.5){\line(1,0){175}}
\multiput(235,259.5)(1,-1){6}{\circle*{.05}}
\put(240,254.5){\line(0,-1){25}}
\put(240,229.5){\line(1,0){25}}

\put(60,229.5){\line(1,0){60}}
\put(120,229.5){\line(0,-1){25}}
\multiput(120,204.5)(1,-1){6}{\circle*{.05}}
\put(125,199.5){\line(1,0){145}}

\put(60,199.5){\line(1,0){55}}
\multiput(115,199.5)(1,-1){6}{\circle*{.05}}
\put(120,194.5){\line(0,-1){25}}
\put(120,169.5){\line(1,0){150}}  }

\put(0,-100){\vector(1,0){370}}

\put(380,-103){$n=\frac{x^{1}}{2}$}

\multiput(43.5,-95)(15,0){17}{\line(0,-1){10}}

\put(0,-100){\vector(0,1){270}}

\put(-4,180){$X=x^{2}$}

\multiput(-5,137.5)(0,-30){7}{\line(1,0){10}}

\put(170,-180){Fig.2}  }

\end{picture}

\newpage


\begin{picture}(400,450)(0,0)

\multiput(10,250)(0,-15){10}{\multiput(0,0)(30,0){5}{\line(1,0){15}}}

\multiput(200,250)(0,-30){5}{\multiput(0,0)(15,0){10}{\line(0,-1){15}}}

\put(70,70){(a)}

\put(260,70){(b)}

\put(160,0){Fig.3}

\end{picture}

\newpage


\begin{picture}(400,450)(0,0)

\multiput(10,250)(30,0){5}{\circle*{1.6}}
\multiput(10,220)(30,0){5}{\circle*{1.6}}
\multiput(10,190)(30,0){5}{\circle*{1.6}}
\multiput(10,160)(30,0){5}{\circle*{1.6}}
\multiput(10,130)(30,0){5}{\circle*{1.6}}

\multiput(0,0)(0,-30){5}{\multiput(0,0)(60,0){3}
{\multiput(10,255)(1,-1){6}{\circle*{.05}}}}

\multiput(0,0)(0,-30){5}{\multiput(0,0)(60,0){3}
{\multiput(5,250)(1,-1){6}{\circle*{.05}}}}

\multiput(0,0)(0,-30){5}{\multiput(15,250)(60,0){2}{\line(1,0){50}}}

\multiput(0,0)(0,-30){4}{\multiput(10,245)(60,0){3}{\line(0,-1){20}}}

\put(70,55){(a)}

\multiput(210,250)(30,0){5}{\circle*{1.6}}
\multiput(210,220)(30,0){5}{\circle*{1.6}}
\multiput(210,190)(30,0){5}{\circle*{1.6}}
\multiput(210,160)(30,0){5}{\circle*{1.6}}
\multiput(210,130)(30,0){5}{\circle*{1.6}}

\multiput(210,250)(0,-30){4}{\multiput(0,0)(60,0){2}{\line(1,-1){30}}}
\multiput(240,250)(0,-30){5}{\multiput(0,0)(60,0){2}{\line(1,0){30}}}

\put(260,55){(b)}

\put(160,0){Fig.4}

\end{picture}

\newpage


\begin{picture}(400,450)(0,0)

\put(60,350){
\multiput(0,0)(0,-30){8}{\multiput(0,0)(30,0){8}{\circle*{1.6}}}}

\put(60,290){\line(1,0){60}}
\put(120,290){\line(0,-1){30}}
\put(120,260){\line(1,-1){30}}
\put(150,230){\line(0,1){30}}
\put(150,260){\line(1,0){60}}
\put(210,260){\line(1,-1){30}}
\put(240,230){\line(1,0){30}}

\put(215,265){{\large $[X]$}}

\multiput(60,260)(3,0){20}{\circle*{.05}}
\multiput(60,230)(3,0){60}{\circle*{.05}}
\multiput(60,200)(3,0){70}{\circle*{.05}}
\multiput(60,170)(3,0){70}{\circle*{.05}}
\multiput(60,140)(3,0){70}{\circle*{.05}}

\multiput(60,170)(3,-3){10}{\circle*{.05}}
\multiput(60,200)(3,-3){20}{\circle*{.05}}
\multiput(60,230)(3,-3){30}{\circle*{.05}}
\multiput(60,260)(3,-3){40}{\circle*{.05}}
\multiput(60,290)(3,-3){50}{\circle*{.05}}
\multiput(90,290)(3,-3){10}{\circle*{.05}}
\multiput(150,230)(3,-3){30}{\circle*{.05}}
\multiput(150,260)(3,-3){40}{\circle*{.05}}
\multiput(180,260)(3,-3){30}{\circle*{.05}}
\multiput(240,230)(3,-3){10}{\circle*{.05}}

\put(120,180){{\large $\cal D$}}

\put(160,30){Fig.5}

\end{picture}

\newpage


\begin{picture}(400,450)(0,0)

\put(60,280){\line(1,0){55}}
\multiput(115,280)(1,-1){6}{\circle*{.05}}
\put(120,275){\line(0,-1){25}}
\put(120,250){\line(1,0){90}}
\put(210,250){\line(1,-1){30}}
\put(240,220){\line(1,0){30}}

\put(30,275){\large $[X]$}

\put(60,310){\line(1,0){60}}
\put(120,310){\line(0,-1){25}}
\multiput(120,285)(1,-1){6}{\circle*{.05}}
\put(125,280){\line(1,0){85}}
\put(210,280){\line(0,-1){25}}
\multiput(210,255)(1,-1){6}{\circle*{.05}}
\put(215,250){\line(1,0){55}}

\put(30,305){\large $[Y]$}

\put(125,285){\bf x}
\put(215,255){\bf z}

\put(60,130){\line(1,0){55}}
\put(125,130){\line(1,0){85}}
\put(210,130){\line(0,-1){30}}
\put(210,100){\line(1,-1){30}}
\put(240,70){\line(1,0){30}}

\put(30,125){\large $[U]$}

\put(60,160){\line(1,0){60}}
\put(120,160){\line(0,-1){60}}
\put(120,100){\line(1,0){85}}
\put(215,100){\line(1,0){55}}

\put(30,155){\large $[V]$}

\put(125,135){\bf x}
\put(215,105){\bf z}

\put(160,30){Fig.6}

\end{picture}

\end{document}